\documentclass[sigconf,screen]{acmart}

\setcopyright{none}
\acmDOI{}
\acmISBN{}
\acmYear{}
\copyrightyear{}
\acmPrice{}
\acmConference[]{}{}{}

\usepackage[linesnumbered,ruled]{algorithm2e}
\usepackage{algorithmicx}
\usepackage{algpseudocode}    
\usepackage{flushend}
\usepackage{amsmath}
\usepackage{epsfig,color}
\usepackage{graphicx}
\usepackage[utf8]{inputenc}   
\usepackage[english]{babel} 
\inputencoding{latin1}
\usepackage{mdwlist}
\usepackage{paralist} 
\usepackage{listings}
\usepackage{multirow}
\usepackage{microtype}
\usepackage{subcaption}
\usepackage{xcolor}
\usepackage{url}
\definecolor{darkblue}{rgb}{0.0, 0.0, 0.55}

\usepackage{tikz}
\usetikzlibrary{positioning}
\usetikzlibrary{calc}
\usetikzlibrary{matrix}

\usepackage{wrapfig}

\usepackage{pgfplots}
\usepackage{pgfplotstable}
\pgfplotsset{compat = 1.3}
\pgfplotsset{
	legend image code/.code={
		\draw[mark repeat=2,mark phase=2]
		plot coordinates {
			(0cm,0cm)
			(0.15cm,0cm)        
			(0.3cm,0cm)         
		};%
	}
}

\usepackage{amsthm}
\theoremstyle{definition}

\let\oldnl\nl
\newcommand{\nonl}{\renewcommand{\nl}{\let\nl\oldnl}}

\newcommand{\ignore}[1]{}

\definecolor{mygreen}{rgb}{0,0.4,0.13}
\definecolor{mygray}{rgb}{0.5,0.5,0.5}
\definecolor{mymauve}{rgb}{0.58,0,0.82}
\definecolor{dkgreen}{rgb}{0,0.5,0}

\algnewcommand\algorithmicforeach{\textbf{for each}}
\algdef{S}[FOR]{ForEach}[1]{\algorithmicforeach\ #1\ \algorithmicdo}

\newcommand{\Name}[1]{\textsf{CANAL}}

\begin{document}

\title{CANAL: A Cache Timing Analysis Framework via LLVM Transformation}

\author{Chungha Sung}
\affiliation{%
  \institution{University of Southern California}
  \city{Los Angeles}
  \state{CA, USA}
}

\author{Brandon Paulsen}
\affiliation{%
  \institution{University of Southern California}
  \city{Los Angeles}
  \state{CA, USA}
}

\author{Chao Wang}
\affiliation{%
  \institution{University of Southern California}
  \city{Los Angeles}
  \state{CA, USA}
}


\begin{abstract}

A unified modeling framework for non-functional properties of a
program is essential for research in software analysis and
verification, since it reduces burdens on individual researchers to
implement new approaches and compare existing approaches.
We present \Name{}, a framework that models the
\emph{cache} behaviors of a program by transforming its
intermediate representation in the LLVM compiler.
\Name{} inserts  auxiliary variables and instructions over these
variables, to allow standard verification tools to handle a new class
of cache related properties, e.g., for computing the worst-case
execution time and detecting side-channel leaks.
%
%
We demonstrate the effectiveness of \Name{} using three verification
tools: KLEE, SMACK and Crab-llvm. We confirm the accuracy of our cache
model by comparing with CPU cycle-accurate simulation results of
GEM5.
\Name{} is available on GitHub\footnote{Tool and benchmarks: \url{https://github.com/canalcache/canal}} 
and YouTube\footnote{Demo video: \url{https://youtu.be/JDou3F1j2nY}}.

\end{abstract}



\maketitle

\section{Introduction}

Analyzing the \emph{cache} behaviors of a program is important, e.g.,
for computing the worst-case execution time of a real-time
system~\cite{li96,chattopadhyay11} and detecting information leaks
through side channels~\cite{DoychevFKMR13,WuGSW18}.
However, existing verification tools are often designed only for
checking \emph{functional} properties, e.g., assertions or pre- and
post-conditions.  For example, none of the participants of recent
software verification competitions~\cite{Beyer15} can
verify \emph{non-functional} properties such as those related to the
execution time.  Although specialized tools have been developed to
handle such non-functional properties, they are rarely open-source or
as well-maintained as mainstream verification tools.
As a result, it is difficult for individual researchers to implement
new approaches for verifying such properties or evaluate existing
approaches.

We fill the gap by developing a \emph{lightweight} cache modeling
framework for standard verification tools, by transforming the LLVM
intermediate representation (IR) of a program to add self-modeling
capabilities. That is, we insert auxiliary variables and LLVM
instructions over these variables to record and update cache
statistics related to \texttt{Load}/\texttt{Store} instructions during
the program execution.  By using the instrumented LLVM bitcode as
input, standard (functional) verification tools will have the
capability of verifying a new class of (non-functional) properties.


Our modeling framework, named \Name{}, takes C/C++ code as input and
emits LLVM bitcode as output.  Thus, it can be used by any LLVM-based
verification tools.
For example, symbolic execution tools such as KLEE~\cite{Cadar08} may
take the program instrumented by \Name{} to detect side-channel leaks;
bounded model checkers such as SMACK~\cite{Rakamaric14} may take the
program instrumented by \Name{} to conduct \textsc{Must-}
and \textsc{May-hit} cache analyses; and static analyzers based on
numerical abstract interpretation, such as Crab-llvm~\cite{Gange16},
may take the program instrumented by \Name{} to conduct worst-case
execution time (WCET) analysis.

In the remainder of this paper, we shall explain how to
combine \Name{} with KLEE, SMACK and Crab-llvm to obtain the desired
results.
We also compare \Name{} with the CPU cycle-accurate simulation results
of GEM5~\cite{Binkert11}, a standard micro-architectural simulator, to
demonstrate the accuracy of our cache model.

\begin{figure}
\hspace{-5ex}
\includegraphics[width=1.1\linewidth]{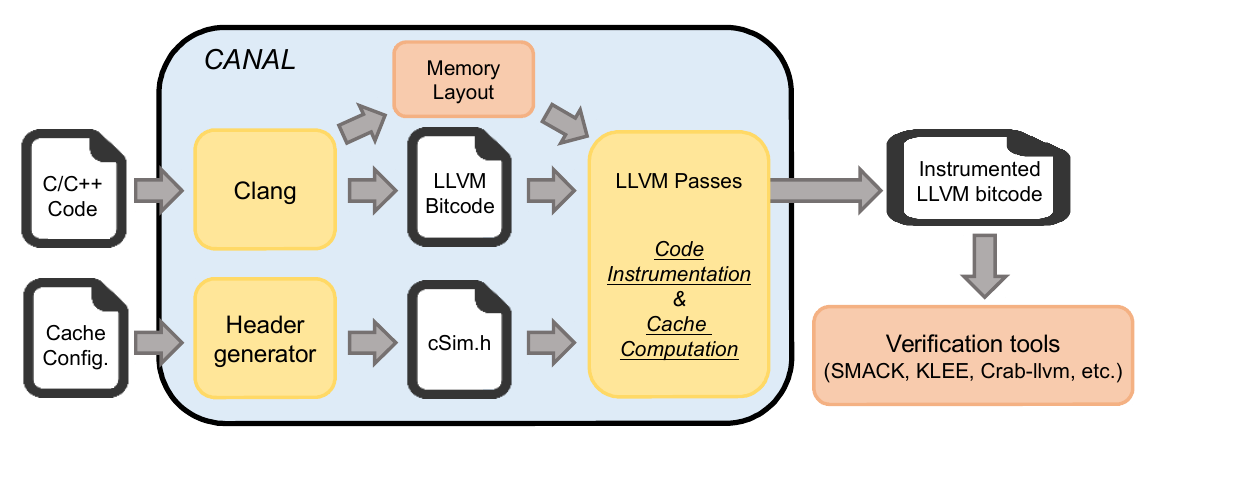}

\vspace{-5ex}
\caption{The overall flow of \Name{}.}
\label{fig:overview}
\vspace{-4ex}
\end{figure}

\section{Tool Overview}

%
Figure~\ref{fig:overview} shows the overall flow of \Name{}, which
takes the C/C++ code of a program and the cache configuration file of
a target computer as input, and returns the instrumented LLVM bitcode
as output.
After compiling the C/C++ code into LLVM bitcode, it uses a sequence
of optimization (\texttt{opt}) passes to insert, before or after
each \texttt{Load}/\texttt{Store} instruction, some new instructions
that model the change of cache states due to these memory accesses.
The inserted instructions can be understood as invocations of two
functions: \texttt{\_\_CSIM\_Load(addrInfo)}
and \texttt{\_\_CSIM\_Store(addrInfo)}, which updates our model of the
cache state whenever \texttt{Load} or \texttt{Store} is
executed; \texttt{addrInfo} denotes information of the memory
location.
%

In addition to the automatically inserted calls
to \texttt{\_\_CSIM\_Load{}} and \texttt{\_\_CSIM\_Store}, the user
of \Name{} may specify properties using these auxiliary variables:
\texttt{\_\_CSIM\_num\_hit}, 
\texttt{\_\_CSIM\_num\_miss}, 
\texttt{\_\_CSIM\_Load\_ret}, and
\texttt{\_\_CSIM\_Store\_ret}.
They represent the accumulative numbers of hits and misses along a
program path, as well as the cache status (hit or miss) associated
with each memory access.
By feeding the instrumented LLVM bitcode to standard verification
tools as input, \Name{} allows them to verify a new class of
non-functional properties, such as assertions over auxiliary variables
that model the cache behaviors of the program.

\begin{wrapfigure}{R}{.45\linewidth}
\centering
\vspace{-3ex}
\begin{minipage}{.97\linewidth}
\begin{lstlisting}
int main() {
	__CSIM_init_cache();
	char out[16];
	aes_encrypt("hello world!", out);
	__CSIM_print_stat();
}
\end{lstlisting}
\end{minipage}
\vspace{-2ex}

\caption{\Name{} as standalone cache simulator.}
\label{fig:usage0}	
\vspace{-2ex}
\end{wrapfigure}

\paragraph{Accuracy}

To demonstrate the accuracy of our cache model, we compare our results
with the cache statistics reported by GEM5.
Toward this end, note that \Name{} may be used as a
standalone cache simulator: if we compile the LLVM bitcode
instrumented by \Name{} to an executable and run it with a concrete
input, it will produce the cache statistics associated with that
particular execution.
Figure~\ref{fig:usage0} shows this usage case,
where \texttt{\_\_CSIM\_init\_cache()} and
\texttt{\_\_CSIM\_print\_stat()} are inserted to the original 
C program to initialize the cache states and display the result,
respectively.  The function body of \texttt{aes\_encrypt()} will be
instrumented by \Name{} automatically.


Since the cache statistics reported by GEM5 include not only
the \texttt{main()} function but also operating system code executed
before and after, we need to create two program versions and then
compute their difference.  One of the programs consists of the
\texttt{main()} function and instructions inserted at the
beginning and end of the \texttt{main()} function to flush the cache,
while the other program consists of only these cache-flushing
instructions with an empty \texttt{main()} function body. By running
these two programs and computing the difference, we have obtained the
exact numbers of cache hits and misses reported by GEM5.

Table~\ref{tbl:accuracy} shows the comparison of GEM5 and \Name{} for
five example programs, including three sorting routines and two
cryptographic routines.  The results are always identical.
The sorting routines exhibit a diverse range of memory-accessing
behavior based on input data (array of random integers).  The
cryptographic routines have security-critical computations that are
often target of side-channel attacks; their inputs are an encrypted
message of ``hello world!''  using a predefined encryption key.
To ensure that we trigger a rich set of cache behaviors during the
experiments, we configured the cache to be 4-way associativity with
LRU replacement policy, 64-byte line size, and 1K-byte cache size.
With a larger cache size, the simulation speed of \Name{} will not
change much but there will be fewer cache conflicts.

\begin{table}[h]
\renewcommand{\arraystretch}{0.6}
\caption{Accuracy comparison: \Name{} versus \textsf{GEM5}}
\label{tbl:accuracy}

\vspace{-1ex}
\centering
\scalebox{0.7}{
\begin{tabular}{lcr|rrr|rrr}
\hline
  &  &  & \multicolumn{3}{c}{\textsf{GEM5}} & \multicolumn{3}{c}{\Name{}} \\
\cline{4-6}\cline{7-9}
Name & LoC & Mem.access
& R.Miss & W.Miss & Time(s) & R.Miss & W.Miss & Time(s) \\
\hline
Ary Acc & 70 & 322,575 
& 2,881 & 0 & 0.62
& 2,881 & 0 & 5.30 \\

Bub.Sort & 49 & 11,028,902 
& 66,560 & 865  & 13.53 
& 66,560 & 865  & 0.32 \\

Ins.Sort & 49 & 2,619,985 
& 15,054 & 68  & 2.98 
& 15,054 & 68  & 0.39  \\

AES~\cite{Aesref} & 789 & 534 
& 171 & 39 & 0.28 
& 171 & 39 & 1.43 \\

DES~\cite{Aesref} & 368 & 580 
& 108 & 11 & 0.29
& 108 & 11 & 0.17 \\

\hline

\end{tabular}
}
\vspace{-2ex}
\end{table}

%

\section{Application Scenarios}

We now demonstrate how \Name{} may be used by KLEE, SMACK, and
Crab-llvm using five example programs taken from the SV-COMP
benchmark~\cite{Beyer15}: \texttt{copysome}, \texttt{sanfoundry},
and \texttt{standard} from the \texttt{array-programs} section,
and \texttt{gcd} and \texttt{sum} from the \texttt{bit-vectors}
section.
We use a 4-way associative cache with LRU and 64-byte cache line while
setting the cache size to 1 KB, 16 KB and 32 KB, respectively.  We set
the timeout to one hour for each program.

\subsection{Combined with Symbolic Execution Tools}

Symbolic execution is a technique for systematically exploring
feasible paths of a program and generating their test inputs.
Although it has been used primarily for checking functional
properties, with \Name{}, it can now be used to detect timing
side-channel leaks.

\paragraph{Timing Side-channel Leaks}

We say that a program $P(k)$ with sensitive input $k$ has timing
side-channel leaks if the execution time of $P$ depends on the value
of $k$.  That is, $\exists k_1,k_2 \in
dom(k): \tau(P,k_1) \neq \tau(P,k_2)$, where $k_1$ and $k_2$ are
values in the domain of $k$ and $\tau$ is the execution time.  Even if
the program executes the same number (and type) of instructions, the
execution time may still differ if there are different numbers of
cache hits/misses.  Such side-channel leaks may be detected by
\Name{} + KLEE.

\begin{wrapfigure}{R}{.45\linewidth}

\centering
\vspace{-3ex}
\begin{minipage}{.97\linewidth}
\begin{lstlisting}
  klee_make_symbolic(&input1);
  klee_make_symbolic(&input2);
  __CSIM_init_cache();
  prog(input1);
  h1 = __CSIM_num_hit;
  m1 = __CSIM_num_miss;
  __CSIM_init_cache();
  prog(input2);
  h2 = __CSIM_num_hit;
  m2 = __CSIM_num_miss;
  assert( h1==h2 && m1==m2 );
\end{lstlisting}
\end{minipage}
\vspace{-2ex}

\caption{\Name{} for timing side channel detection.}
\label{fig:usage1}	
\vspace{-2ex}
\end{wrapfigure}

Figure~\ref{fig:usage1} shows an example, where \texttt{input1}
and \texttt{input2} are marked as symbolic values and used to run the
program \texttt{prog()} twice.  After each execution, the numbers of
hits and misses are stored in \texttt{h1}, \texttt{m1}, \texttt{h2},
and \texttt{m2}, respectively.  Finally, the assertion checks
if \texttt{prog()} is leak-free; that is,
$\forall$ \texttt{input1,input2}, the condition (\texttt{h1==h2 \&\& m1==m2}) always holds.
KLEE can be used to search for concrete values of \texttt{input1}
and \texttt{input2} that violate the assertion.

Table~\ref{tbl:result1} shows the results of running KLEE on these
programs.  In each case, we manually modified the program to mark one
or more parameters as the sensitive input. Columns 3-5 show if a leak
is detected, together with the total number of tests generated and,
among them, the number tests that manifest the leak.  Columns 6-8 show
the time taken by KLEE for the different cache sizes.

\begin{table}[h]
\renewcommand{\arraystretch}{0.7}
\caption{Results of the side-channel leak detection.}
\label{tbl:result1}

\vspace{-1ex}
\centering
\scalebox{0.7}{
\begin{tabular}{lc|ccc|rrr}
\hline
           &      & \multicolumn{3}{c|}{\Name{} + KLEE} & \multicolumn{3}{c}{Time (s)} \\
\hline
Name       & LoC  & Detection result   & Total-tests     & Leaky-tests        & 1 KB  & 16 KB  & 32 KB \\
\hline
copysome   & 79   & No leak     &  121        &  0             & 1.38 & 5.05  & 13.19 \\ 
sanfoundry & 95   & No leak     &   81        &  0             & 1.32 & 2.16  & 2.82  \\ 
standard   & 61   & Leak        &    9        &  1             & 0.27 & 0.69  & 1.10  \\ 
gcd        & 52   & Leak        &    6        &  1             & 0.11 & 0.41  & 0.77  \\ 
sum        & 56   & Leak        &    6        &  1             & 0.07 & 0.38  & 0.74  \\ 
\hline

\end{tabular}
}
\vspace{-2ex}
\end{table}

\subsection{Combined with Software Verification Tools}

While symbolic execution is geared toward generating tests, verification tools such as SMACK are geared toward generating proofs,
e.g., proving that an assertion holds under all test inputs.
We show how SMACK can leverage \Name{} to prove cache related
properties.

\paragraph{Must-hits and Must-misses}

One type of properties of interest is assertions over auxiliary
variables such as \texttt{\_\_CSIM\_Load\_ret}
and \texttt{\_\_CSIM\_Store\_ret}.
For example, if a certain \texttt{Load} or \texttt{Store} instruction
in the program always leads to a cache hit or miss, regardless of the
program path and test input; in such a case, we call it a Must-hit or
a Must-miss.

\Name{} instrument the LLVM bit-code in such a way that calls
to \texttt{\_\_CSIM\_Load()} and \texttt{\_\_CSIM\_Store()} set the
values of auxiliary variables \texttt{\_\_CSIM\_Load\_ret}
and \texttt{\_\_CSIM\_Store\_ret} to reflect the cache
status: \emph{true} means the memory access leads to a hit,
whereas \emph{false} means it leads to a miss.

Figure~\ref{fig:usage2} shows a program where we check if read
of \texttt{buffer[2]} is a Must-hit.  Thus, we save the value
of \texttt{\_\_CSIM\_Load\_ret} immediately after the read
of \texttt{buffer[2]} to the variable named \texttt{h} and add an
assertion stating \texttt{h} should always be \emph{true}.
If SMACK can prove the assertion, we know the read
of \texttt{buffer[2]} is a Must-hit.  Alternatively, we can
add \texttt{assert(h==false)} and use SMACK to prove it is a
Must-miss.  

Since program verification is undecided in general (e.g., equivalent
to the Turing-halting problem), SMACK may fail to prove either
assertion; in such a case, the result remains inconclusive.
In this particular example, however, SMACK is able to find a violation
of the Must-hit assertion and generate a counterexample.  The
counterexample shows a scenario where \texttt{buffer[0]}
and \texttt{buffer[16]} resides in two different 64-byte cache lines.


\begin{wrapfigure}{R}{.45\linewidth}
\centering
\vspace{-3ex}
\begin{minipage}{.97\linewidth}

\begin{lstlisting}
  if (cond)  buffer[0] = x;
  else       buffer[16] = y;
  z = buffer[2];
  h = __CSIM_Load_ret;
  assert(h == true); // `Must-Hit'?
\end{lstlisting}
\end{minipage}
\vspace{-2ex}

\caption{\Name{} for must-hit analysis.}
\label{fig:usage2}	
\vspace{-2ex}
\end{wrapfigure}

Table~\ref{tbl:result2} shows the results of applying \Name{}+SMACK to
assertions we manually inserted to check if a \texttt{Load}
or \texttt{Store} instruction in the program is a \textsc{Must}-hit/miss.
With loop-unrolling bound of SMACK set to 10, and the cache size set
to 1 KB, SMACK successfully verified all assertions.
However, when the cache size was increased to 16 KB and 32 KB, SMACK
started to timeout on some programs. This points out a scalability
limitation of SMACK, together with direction for future work:
improving the verification algorithms to make SMACK (and similar
tools) more scalable for non-functional properties.

\begin{table}[h]
\renewcommand{\arraystretch}{0.7}
\caption{Results of the Must-hit analysis.}
\label{tbl:result2}

\vspace{-1ex}
\centering
\scalebox{0.7}{
\begin{tabular}{lc|ccc|rrrr}
\hline
&  & \multicolumn{3}{c|}{\Name{} + SMACK} & \multicolumn{3}{c}{Time (s)} \\
\hline
Name & LoC & Loop-unroll-bound & Property & Results & 1 KB & 16 KB & 32 KB \\
\hline
copysome   & 69 & 10 & Must-miss  & Verified & 226.87 & TO      & TO \\ 
sanfoundry & 75 & 10 & Must-miss  & Verified &  78.08 & 1055.55 & TO \\ 
standard   & 38 & 10 & Must-hit   & Verified &  24.53 & 139.59  & 344.71 \\ 
gcd        & 74 & 10 & Must-miss  & Verified &  22.88 & 142.78  & 375.27 \\ 
sum        & 42 & 10 & Must-miss  & Verified &  36.53 & 257.68  & 723.21 \\

\hline

\end{tabular}
}
\vspace{-2ex}
\end{table}

\subsection{Combined with Static Analysis Tools}

Static analyzers based on numerical abstract
interpretation~\cite{CousotC77}, such as Crab-llvm, can generate
program invariants.  These invariants, computed for each program
location, are summaries over all paths and input values.  Therefore,
they can be used to estimate the worst-case execution time of a
program.  More specifically, by leveraging \Name{}, tools such as
Crab-llvm can generate invariants in terms of auxiliary variables such
as ($5\leq \_\_CSIM\_num\_Load\_hit \leq 18$).

\begin{wrapfigure}{R}{.45\linewidth}
\centering
\vspace{-3ex}
\begin{minipage}{.97\linewidth}
\begin{lstlisting}
 if (cond)  buffer[0] = 1;
 else       buffer[16] = 1;
 buffer[5] = 1;
 n_s_h = __CSIM_num_Store_hit;
 n_s_m = __CSIM_num_Store_miss;
 n_s = n_s_h + n_s_m;
 __CRAB_assert(n_s_h  > 1);
 __CRAB_assert(n_s_m < 3);
 __CRAB_assert(n_s == 2);
\end{lstlisting}
\end{minipage}
\vspace{-2ex}

\caption{\Name{} for computing value ranges.}
\label{fig:usage3}	
\vspace{-2ex}
\end{wrapfigure}

Figure~\ref{fig:usage3} shows an \texttt{if-else} statement controlled
by the value of \texttt{cond}.  Assume each cache line contains 64
bytes, the first 16 integers of the array fall into one cache line,
whereas the next 16 integers starting with
\texttt{buffer[16]} fall into another cache
line.  However, during static analysis, there is no way of knowing
what the value of \texttt{cond} is; therefore, one has to assume both
branches may be taken.

When the \emph{Then}-branch is taken, \texttt{buffer[5]} will be
loaded to the cache, which means the access to \texttt{buffer[5]} is a
cache hit.  However, when the \emph{Else}-branch is
taken, \texttt{buffer[5]} will not be loaded to the cache, which means
the access to \texttt{buffer[5]} is a cache miss.  By using numerical
abstract interpretation, Crab-llvm can take both cases into
consideration and compute value ranges
of \texttt{n\_s}, \texttt{n\_s\_h} and \texttt{n\_s\_m}.
For this example, in particular, the value ranges would be [2,2]
for \texttt{n\_s}, [0,1] for \texttt{n\_s\_h}, and [1,2]
for \texttt{n\_s\_m}.  Therefore, Crab-llvm can prove the second and
the third assertions, while reporting a \emph{potential} violation of
the first assertion.

In addition, an interesting application of the value ranges computed
by numerical abstract interpretation is to compute the worst-case
execution time (WCET), which depends on the maximum number of cache
misses along all program paths.

\begin{table}[h]
\renewcommand{\arraystretch}{0.7}
\caption{Results of the numerical abstract interpretation.}
\label{tbl:result3}

\vspace{-1ex}
\centering
\scalebox{0.685}{
\begin{tabular}{lc|rrrr|rrrr}
\hline

& & \multicolumn{4}{c|}{\Name{} + Crab-llvm}  & \multicolumn{3}{c}{Time (s)}  \\
\hline

Name & LoC & S.Hit & S.Miss & L.Hit & L.Miss & 1 KB & 16 KB & 32 KB \\
\hline
copysome & 75 & [1, $\infty$] & [2, $\infty$] & [1, $\infty$] & [0, $\infty$] & 73.77 & 937.23 & TO\\
sanfoundry & 85 & [0, $\infty$] & [3, $\infty$] & [4, $\infty$] & [1, $\infty$] & 67.17 & 636.89 & 2416.82 \\
standard & 58 & [0, $\infty$] & [1, $\infty$] & [1, $\infty$] & [0, 0] & 13.06 & 528.53 & 2188.98 \\
gcd & 82 & [0, $\infty$] & [2, $\infty$] & [6, $\infty$] & [0, $\infty$] & 3.99 & 105.59  & 382.53 \\
sum & 54 & [0, $\infty$] & [3, 3] & [2, $\infty$] & [0, 0] & 0.87 & 39.03 & 146.01 \\
\hline

copysome-unroll & 105 & [22, 22] & [4, 4] & [43, 43] & [0, 0] & 91.62 & 452.34 & 1303.80\\
sanfoundry-unroll & 168 & [5, 15]  & [3, 3]  & [14, 29]  & [1, 5]  & 19.04 & 296.64 & 1149.54 \\
standard-unroll & 130 & [10, 19] & [2, 11] & [20, 20] & [0, 0] & 21.11 & 279.30 & 985.24 \\
gcd-unroll & 107 & [0, 13] & [2, 3] & [6, 27] & [0, 0] & 11.69 & 174.16 & 608.06 \\
sum-unroll & 123 & [6, 12] & [3, 3] & [20, 32] & [0, 0] & 12.51 & 197.88 & 688.32 \\
\hline

\end{tabular}
}
\vspace{-2ex}
\end{table}

Table~\ref{tbl:result3} shows the results of
applying \Name{}+Crab-llvm on the example programs.  Columns 3-6
report the value ranges of the total number Store-hits (S.Hit),
Store-misses (S.Miss), Load-hits (L.Hit) and Load-misses (L.Miss).
Since these programs have loops and Crab-llvm uses aggressive
over-approximation to force termination over loops, most of the upper
bounds become $+\infty$.
Luckily, these are fixed-bound loops, and after we automatically
unrolled these loops, Crab-llvm obtained more accurate value ranges.

\section{Cache Modeling}

We now briefly explain how cache is modeled inside \Name{}. It is a 
\emph{lightweight} cache model in that the modeling instructions 
are carefully designed to reduce the overhead of the verification
tools.  For example, pointers are difficult to handle by verification
tools; therefore, we avoid using them in the instrumented code.

\subsection{Pre-computing Address-to-Cache Mapping}

Inside LLVM, we first obtain the memory address of each program
variable by analyzing the symbol table of the pre-compiled code.
Then, for the target computer architecture, we generate a memory
layout.  We try to pre-compute the possible address value for each
load or store instruction in the program.
If the address is a fixed value, we compute its \emph{set}
and \emph{tag} fields in the cache, and use these concrete values to
simplify the instantiation of \texttt{\_\_CSIM\_Load()}
and \texttt{\_\_CSIM\_Store()}.  Otherwise, we resort to the use of
if-else statements to dynamically compute the \emph{set}
and \emph{tag} fields (more difficult to handle by
verification tools).

\begin{wrapfigure}{R}{.45\linewidth}
\vspace{-3ex}
\begin{minipage}{.97\linewidth}
\begin{lstlisting}
int var;//its cache `set' and `tag' are 242 and 1, respectively
var = 2;
__CSIM_Store(242, 1);
\end{lstlisting}
\end{minipage}
\vspace{-2ex}
\caption{Pre-computed `set' and `tag' values.}
\label{fig:precompute1}	
\vspace{-2ex}
\end{wrapfigure}

Figure~\ref{fig:precompute1} shows a simple case where the address
of \texttt{var} is statically known, and thus we can pre-compute
its \emph{set} (242) and \emph{tag} (1).  These concrete values are
used to instantiate \texttt{\_\_CSIM\_Store()}. Although C code is
used in Figure~\ref{fig:precompute1}, this is actually implemented at
the LLVM bitcode level inside \Name{}.

\begin{wrapfigure}{R}{.5\linewidth}
\vspace{-3ex}
\begin{minipage}{.97\linewidth}
\begin{lstlisting}
buffer[i] = 20;
if (address_of_buffer + 4*i < __CSIM_addr_of_cacheline01) {
		__CSIM_Store(242, 0);
} else if (address_of_test + 4*i < __CSIM_addr_of_cacheline02) {
   		__CSIM_Store(242, 1);
} else {
  		__CSIM_Store(242, 3);
}
\end{lstlisting}
\end{minipage}

\vspace{-2ex}
\caption{Dynamically computed `set' and `tag' values.}
\label{fig:precompute2}	
\vspace{-2ex}
\end{wrapfigure}

Figure~\ref{fig:precompute2} shows a more complex case, where the
array is accessed using a variable $i$, and thus if-else statements
are used to compute the \emph{set} and \emph{tag}
of \texttt{buffer[i]}.
Although we cannot simplify as much as in
Figure~\ref{fig:precompute1}, we can still pre-compute the value
ranges of \emph{set} and \emph{tag} based on the address of the array.
In particular, we can assume the value range of \emph{set} is [0,3]
and the \emph{tag} is always 242.

\subsection{Simplifying Updates of the Cache Statistics}

To simplify the storage and update of cache statistics so verification
tools can handle them easily, we use a set of simple variables as
opposed to an array indexed by memory addresses.  This can drastically
reduce the complexity of the cache-modeling instructions inside
functions \texttt{\_\_CSIM\_Load()} and \texttt{\_\_CSIM\_Store()}.

\begin{wrapfigure}{R}{.47\linewidth}
\vspace{-1ex}
\begin{minipage}{.99\linewidth}
\begin{lstlisting}
function __CSIM_Store(set, tag) {
  if (set == 0) {
    if (__CSIM_cacheline00_taken && 
        __CSIM_cacheline00_tag==tag){  
       // cache hit
       __CSIM_num_Store_hit ++;
       __CSIM_Store_ret = true;
    } else if (...) {
       // cache hit
       ...
    } else { 
       // cache miss
       __CSIM_num_Store_miss ++;
       __CSIM_Store_ret = false;
       // pick a new line based on the update policy
       ...
    }
  } else if (set == 1) {
    ...
  } else if (set == 2) {
    ...
  }
  ...
}   	
\end{lstlisting}
\end{minipage}
\vspace{-2ex}

\caption{Code snippet of \texttt{\_\_CSIM\_Store}.}
\label{fig:cacheModel}	
\vspace{-2ex}
\end{wrapfigure}

Figure~\ref{fig:cacheModel} shows the internals
of \texttt{\_\_CSIM\_Store()}, which updates the cache statistics
based on the values of \emph{set} and \emph{tag}.  Instead of using
monolithic arrays such as \texttt{cacheline[set].tag}, we use
individual variables such as \texttt{\_\_CSIM\_cacheline00\_tag}.

The number 00 means the cache line is associated with set 0 and way 0,
and the auxiliary variable denotes the tag saved at the line.  When a
cache miss occurs, for example, we update the value
of \texttt{cacheline00\_tag} as well as the values of similar
auxiliary variables, and evicts a victim.  In this implementation, LRU
policy is used to compute the victim; but other replacement policies
may be incorporated into \Name{} easily.

Implementations of functions \texttt{\_\_CSIM\_Store()}
and \texttt{\_\_CSIM\_Load()} are specific to each individual program
under verification, and therefore they are generated by \Name{}
automatically.

\section{Related Work}

\Name{} is the first LLVM-based lightweight cache modeling framework 
designed specifically for software verification tools.  Although there
are other cache simulators~\cite{Cabeza99,jaleel08} and CPU simulators
such as GEM5~\cite{Binkert11}, they are not designed for this
purpose. In particular, they cannot be used in the same way as \Name{}
to afford existing verification tools the capability of verifying a
new class of cache related non-functional properties.

There are also tools designed specifically for WCET analysis based on
cache analysis~\cite{li96,chattopadhyay11,chu16} and for detecting
cache timing side channels~\cite{DoychevFKMR13,WuGSW18,GuoWW18}. However, the
modeling part of these tools are tied up with the subsequent analysis
part, and therefore cannot be used by other verification tools.
Furthermore, the analysis part of these tools is rarely open-source,
and often not as well-maintained as the mainstream software
verification tools, which are updated constantly to keep up with the
competition~\cite{Beyer15}.

Although our main contribution in this work is the lightweight cache
modeling that facilitates the subsequent analysis and verification,
there is still room for improvement in the analysis and verification
algorithms.  Since cache timing behaviors are \emph{non-functional}
properties, they often have significantly different characteristics
from \emph{functional} properties, and thus may benefit from
specialized algorithms to make verification more efficient and
scalable.


Our implementation of \Name{} has been tested on programs from two
domains: real-time software and embedded software.  In both cases, the
program structure and language constructs are relatively simple.  To
handle C/C++ programs in other application domains, more sophisticated
static analyses may be needed, e.g., to deal with pointer aliasing and
complex loops during the pre-computation of address-to-cache mapping
and updates of the cache statistics, in order to keep the application
of our LLVM based transformation efficient.
We also plan to further refine our cache model, e.g., to handle
multi-threading as well as multi-level cache.

\section{Conclusions}
\label{sec:conclusion}

We have presented \Name{}, a framework for modeling cache behaviors of
a program based on LLVM transformations.  \Name{} allows standard
software verification tools to check a new class of cache timing
related properties.  We have demonstrated the accuracy of our cache
model in \Name{} by comparing with the simulation results
of \textsf{GEM5}, as well as the effectiveness of combining \Name{}
with three existing tools (KLEE, SMACK and Crab-llvm) in verifying
cache related properties.

\bibliographystyle{plain}
\bibliography{cacheSim}

\end{document}